# Learning Optimal Decoherence Time Formulas for Surface Hopping Simulation of High-Dimensional Scattering

*Cancan Shao, Rixin Xie, Zhecun Shi, and Linjun Wang**

School of Intelligent Manufacturing, Zhejiang Polytechnic University of Mechanical

and Electrical Engineering, Hangzhou 310053, China

Zhejiang Key Laboratory of Excited-State Energy Conversion and Energy Storage,

Department of Chemistry, Zhejiang University, Hangzhou 310058, China

*Email: ljwang@zju.edu.cn

**ABSTRACT:** In our recent work (*J. Phys. Chem. Lett.* **2023**, *14*, 7680), we utilized the exact quantum dynamics results as references and proposed a general machine learning method to obtain the optimal decoherence time formula for surface hopping simulation. Here, we extend this strategy from one-dimensional systems to the much more intricate scenarios with multiple nuclear dimensions. Different from the one-dimensional situation, an effective nuclear kinetic energy is defined by extracting the component of nuclear momenta along the non-adiabatic coupling vector. Combined with the energy difference between adiabatic states, high-order descriptor space can be generated by binary operations. Then the optimal decoherence time formula can be obtained by machine learning procedures based on the full quantum dynamics reference data. Although we only use the final channel populations in 24 scattering samples as training data for machine learning, the obtained optimal decoherence time formula can well reproduce the time evolution of the reduced and spatial distribution of population. As benchmarked in a large number of 56840 one- and two-dimensional samples, the optimal decoherence time formula shows exceptionally high and uniform performance when compared with all other available formulas.

Reliable and efficient non-adiabatic dynamics simulation methods are essential for the understanding of many complex phenomena in chemistry and materials science,[1–7] such as interfacial charge transfer,[1] photocatalytic reduction,[4,5] and polariton relaxation,[7] etc. To this end, numerous mixed quantum-classical dynamics methods, which treat electrons and nuclei as quantum and classical particles respectively, have been proposed.[8–16] In particular, Tully's fewest switches surface hopping (FSSH)[9] and related variations[17–23] have been widely utilized due to their ease of implementation and acceptable accuracy. The independent propagation of surface hopping trajectories efficiently yields statistical properties, which can be regarded as a measurement process for the mixed quantum-classical system.

In quantum mechanics, coherence refers to the property that allows quantum states to exist in superposition, meaning they can be in multiple states simultaneously. Due to interactions with the environment or other factors, this coherence can degrade over time, leading to a loss of the quantum superposition and a transition towards classical behavior. Nevertheless, tackling the coupling between electrons and nuclei and accurately delineating electronic coherence and decoherence are still challenging for conventional mixed quantum-classical dynamics methods[24–32] such as FSSH, which encounter limitations. Thus far, many decoherence algorithms have been proposed to address this challenge, including the direct collapse of the wave function to the active state when classical trajectories deviate from the non-adiabatic coupling (NAC) region,[33,34] the instantaneous resetting of wave function coefficients subsequent to surface hops,[35] and the classification adjustments to the wave function with respect to the judgement of trajectory branching,[19,21,23] among other considerations.

Other than the treatments mentioned above, the decay of coherence can be interpreted as a rate process that can be depicted by a general time-dependent rate coefficient. From a numerical perspective, during each time step, the wave function coefficient $w_i$ for the *i*-th nonactive state can be corrected as

$$w'_{i(\neq a)} = w_i \exp(-\Delta t / \tau_{ai}), \tag{1}$$

where $\Delta t$ is the time-step size and $\tau_{ai}$ is the decoherence time. The coefficient of the active state *a* is then reset through

$$w'_a = \frac{w_a}{|w_a|}\sqrt{1 - \sum_{i \neq a}|w'_i|^2}. \tag{2}$$

In the literature, different decoherence time formulas have been proposed. According to the quantum wave packet (WP) picture, decoherence process can be interpreted as the decay of overlap between vibronic states. As the motion of WPs can be approximately by semiclassical trajectory ensembles, the decoherence process can be related to the trajectory's momenta, forces or energies. For instance, Rossky and colleagues introduced the force-based decoherence time formula,[24,36,37] which is based on the frozen Gaussian wave packet approximation.[38,39] Subotnik and Shenvi formulated a more rigorous evaluation of the decoherence rate using the quantum Liouville equation and put forward the augmented FSSH algorithm to incorporate proper decoherence.[26,40–42]

Actually, the primary considerations revolve around the convenience and efficiency of decoherence operation in practical applications. Truhlar and co-workers proposed a series of "decay-of-mixing" methods[43–46] and derived a general expression for the decoherence time,

$$\tau_{ai} = \frac{\hbar}{|E_a - E_i|}\left(1 + \frac{A}{(\mathbf{P}\cdot\hat{s})^2/2\mu}\right). \tag{3}$$

Here, $\hbar$ denotes the reduced Planck constant, $|E_a - E_i|$ signifies the absolute energy difference between the active state and the nonactive state, $\mu$ represents nuclear mass, and the momentum $\mathbf{P}$ is directed along the decoherent $\hat{s}$, which is taken along the NAC vector in regions of strong interaction and along the vibrational momentum elsewhere. Later, a more concise energy-difference based decoherence time formula[47] is commonly employed when simulating non-adiabatic dynamic processes in realistic systems[48–51] which serves as

$$\tau_{ai} = \frac{\hbar}{|E_a - E_i|}\left(1 + \frac{A}{E_{kin}}\right), \tag{4}$$

where $E_{kin}$ stands for the overall nuclear kinetic energy, and $A$ is an empirical parameter typically assigned a value of 0.1 au for both eqs 3 and 4. Atomic units are used unless otherwise noted.

The decoherence time in eq 4 decreases as $E_{kin}$ increases and approaches the limit of

$$\lim_{E_{kin} \to \infty} \tau_{ai} = \frac{\hbar}{|E_a - E_i|}. \tag{5}$$

However, the corresponding decoherence rate should be infinitely small when the kinetic energy approaches infinite, as WPs on different adiabatic potential energy surfaces (PESs) propagate in a comparable manner in this case, which results in an infinitely slow decay of the overlap between vibronic states. Based on this comprehension, we have introduced other decoherence time formulas where $E_{kin}$ transitions from the denominator to the numerator,[52] including the linear equation

$$\tau_{ai} = \frac{\hbar}{|E_a - E_i|}(A + BE_{kin}) \tag{6}$$

and the exponential form

$$\tau_{ai} = A \exp\left(\frac{BE_{kin}}{|E_a - E_i|}\right) \tag{7}$$

where $A = 0.5$, $B = 1200$ for eq 6, and $A = 25$, $B = 42$ for eq 7. As benchmarked in numerous one-dimensional two-level scattering models, surface hopping methods with these two modified formulas yield superior results compared to the widely used decoherence time formula, eq 4. Moreover, there remains significant opportunity for further enhancements. For example, these two eqs 6 and 7 fail in the low-energy region of some complex models, and are inferior to the existing branching corrected surface hopping (BCSH) method when treating multilevel systems.[53]

Rather than acquiring individual formulas from physical pictures or mathematical derivations, a unified methodology for constructing decoherence time formulas could significantly benefit from integrating a machine learning framework. The decoherence time formula can be expressed through the Taylor expansion as

$$\tau_{ai} = C_0 + C_1 f + C_2 f^2 + ... + C_n f^n \tag{8}$$

where $\{C_i\}$ represent the expansion coefficients attainable through a discrete optimization algorithm, and $f$ denotes descriptors that can be symbolically expressed by the essential dynamical variables. For the sake of brevity, we can begin with a primary descriptor space ($\Phi_0 = [\delta_1, \delta_2]$), where $\delta_1$ and $\delta_2$ represent two predefined descriptors. More intricate descriptors for $f$ can be derived by applying binary operations $o_i(\cdot,\cdot)$ to the original descriptors in $\Phi_0$, resulting in a new descriptor space ($\Phi_1 = [o_1(\delta_1,\delta_2), o_2(\delta_1,\delta_2), \cdots, o_M(\delta_1,\delta_2)]$). Ultimately, we can construct a higher-order descriptor space through iterative algebraic operations, wherein the complexity of $f$ is governed by truncating the number of Taylor expansion items $n$ in eq 8.

As an initial attempt in a previous study,[53] we focused on the first-order expression of eq 8, i.e., $\tau_{ai} = C_0 + C_1 f$. The primary descriptor space was defined as $\Phi_0 = [E_{kin}, \Delta E_i]$, where $\Delta E_i = E_i - E_a$ can distinguish between the energy levels of two non-active states, even if they have the same absolute energy difference relative to the active state. Then we obtained the first-order descriptor space $\Phi_1$ by applying nine selected binary operations to $\Phi_0$, and further produced $\Phi_2$ which involves the same binary operations for all the descriptors in $\Phi_1$ and $\Phi_0$. After multilayer screening with prepared models selected from the two-level model base, five descriptors with outstanding performance were prominent among 371 descriptors in $\Phi_2$. A representative one was

$$\tau_{ai} = C_0 + C_1 \frac{\max(E_{kin}, \Delta E_i)}{E_{kin} - \Delta E_i} \tag{9}$$

where $C_0$ = 2.5×10$^{-3}$ and $C_1$ = 2.0×10$^5$. In the subsequent evaluation of diverse multilevel systems and six standard scattering models, surface hopping with eq 9 yielded accurate results comparable to those of the BCSH method. Furthermore, this formulaic approach, which learns directly from dynamic data, validates the accuracy of the BCSH method which corresponds to the physical representation of WP reflection.

In a series of one-dimensional multilevel systems, it is encouraging to observe that the new decoherence time formula, such as eq 9, produce by the new machine learning-aided methodology can efficiently reproduce the exact results of quantum dynamics as the conventional eq 4. Nonetheless, the more formidable challenge lies in high-dimensional systems, and we anticipate deriving appropriate decoherence descriptors through an analogous strategy. A schematic diagram of screening processes for the machine learning-aided decoherence time formula approaches is illustrated in Figure 1, where the generation of the training dataset and descriptor space is essential. Given the insufficient depth of research into complex high-dimensional systems, the study should begin with constructing non-trivial high-dimensional models and obtaining the reference quantum dynamics results. In prior academic studies, Subotnik and colleagues introduced several two-dimensional scattering models.[40,41,54] Building upon this groundwork, we can generate a broad range of training data within the STD models by considering different initial conditions for dynamical simulations.

We first consider the STD-1 model proposed by Subotnik.[41] The Hamiltonian reads

$$H_{11}(x, y) = -A_1 \tanh(B_1 x), \tag{10}$$

$$H_{22}(x, y) = A_2 \tanh[B_2(x-1) + C_2 \cos(D_2 y + \pi/2)] + 3A_2/4, \tag{11}$$

$$H_{12}(x,y) = H_{21}(x,y) = A_3 \exp(-B_3 x^2), \tag{12}$$

where $A_1 = 0.05$, $B_1 = 0.6$, $A_2 = 0.2$, $B_2 = 0.6$, $C_2 = 2.0$, $D_2 = 0.3$, $A_3 = 0.015$, and $B_3 = 0.3$. We place the initial Gaussian WP on either the lower or upper surface at coordinates $(x_0, y_0)$, where $x_0 = -4.0$ and $y_0$ is chosen from $\{-2.0, -1.0, 0, 1.0, 2.0, 3.0, 4.0, 5.0\}$. The initial momentum is directed at an angle ($\theta_0$) chosen from $\{0°, 15°, 30°, 45°\}$ relative to the $x$-axis. For scenarios where the initial WP is situated on the lower (upper) surface, the momenta span from 16~28 (8~20). Therefore, we have considered 832 distinct initial conditions. We confine the dynamics within a square region defined by $-15 \leq x, y \leq 25$, and the simulation of exact quantum dynamics ends once the electronic population exceeds a certain threshold at the boundaries.

We also consider the more complex STD-2 model, where the two dimensions are entangled strongly.[54] The corresponding diabatic Hamiltonian is given by

$$H_{11}(x,y) = -E_0, \tag{13}$$

$$H_{22}(x,y) = -A \exp\{-B[0.75(x+y)^2 + 0.25(x-y)^2]\}, \tag{14}$$

$$H_{12}(x,y) = H_{21}(x,y) = C \exp\{-D[0.25(x+y)^2 + 0.75(x-y)^2]\}, \tag{15}$$

where $A = 0.15$, $B = 0.14$, $C = 0.015$, $D = 0.06$, and $E_0 = 0.05$. To perform a comprehensive study, for the initial WP, $x_0$ is set as $-8$ and $y_0$ ranges from $-4$~1, while the angle relative to the $x$-axis $\theta_0$ is set as $0°$, $15°$, or $30°$. For trajectories starting on the lower (upper) surface, the initial WP momenta span from 8~20 (or 4~16). As a result, we can examine channel populations for a total of 468 initial conditions. The dynamics are confined to a square region defined by $-15 \leq x, y \leq 15$.

The Hamiltonian of the STD-3 model is defined as[40]

$$H_{11}(x,y) = A\tanh[B(x+7)] + 2A\tanh[Bx + 2.2\cos(y+1.33)] + 2A, \quad (16)$$

$$H_{22}(x,y) = -A\tanh[B(x+7)], \quad (17)$$

$$H_{12}(x,y) = H_{21}(x,y) = C\exp[-(x+7)^2][2 + \cos(0.8y + 0.46)]. \quad (18)$$

Here, $A = 0.03$, $B = 1.6$, and $C = 0.004$. We place the initial Gaussian WP on either the lower or upper surface at coordinates $(x_0, y_0)$. Here, $x_0$ is set as -12.0, while $y_0$ varies from -5.0 to 0.0. The initial momentum is oriented at angles of $\{0°, 15°, 30°, 45°\}$ with respect to the x-axis. For WPs initialized on the lower (upper) surface, the momenta range from 14~22 (8~16), respectively. Therefore, there are 432 distinct initial conditions. The dynamics are confined within a square region defined by $-15 \leq x \leq 20$ and $-15 \leq y \leq 25$.

For the three different two-dimensional models discussed above, the spatial width of the initial WP is uniformly set to $\sigma_x = \sigma_y = 0.5$ and $\sigma_{px} = \sigma_{py} = 1$. The dynamics pathways may be significantly different for the 1,732 different initial conditions. For convenience, we employ two boundary lines (i.e., $x = 5.0$ and $y = 5.0$ in the STD-1 model, $x = 0.0$ and $y = 0.0$ in the STD-2 model, and $x = 2.5$ and $y = 5.0$ in the STD-3 model) to uniformly divide the entire space into eight channels. Four of them correspond to the lower PES, while the other four are on the upper PES. For each initial condition with the parameters $(y_0, k_0, \theta_0)$, the population error of the $i$th sample is defined as

$$\varepsilon_i = \sqrt{\frac{1}{N_{channel}} \sum_{j=1}^{N_{channel}} \left(P_{ij}^{SH} - P_{ij}^{DVR}\right)^2}. \quad (19)$$

Here, $N_{channel}$ denotes the number of channels for the two-dimensional samples and is set to 8. $P_{ij}^{SH}$ and $P_{ij}^{DVR}$ are populations of the $j$th channel in the $i$th sample at the final snapshot, calculated by surface hopping (SH) and the discrete variable representation (DVR), respectively.

In Figure 2, we show the error distribution of the samples in STD models calculated by FSSH and BCSH. It is worth noting that the performance of BCSH can serve as a crucial reference to show different mechanism of decoherence. For instance, a large error in BCSH results indicates that the rapid WP separation mechanism may not dominate the decoherence process. In the STD-1 model, FSSH yields a markedly larger number of samples with errors exceeding 0.02, whereas BCSH effectively reduces the errors of those samples (see Figures 2A and 2B). In the STD-2 model, however, BCSH with decoherence correction still yields significant errors (see Figure 2D). For the STD-3 model, FSSH has already demonstrated high performance, and BCSH consistently holds the accuracy (see Figures 2E and 2F). In summary, the three STD models encompass three representative cases, which constitute a reasonable training data set for further machine learning.

As BCSH performs well in STD-1 and STD-3 models, the samples exhibiting large errors calculated by BCSH in the STD-2 model become particularly important. This indicates that the dynamics of these samples may involve decoherence mechanisms that differ significantly from those observed in one-dimensional scattering systems. Thereby, we focus on the samples from the STD-2 model for machine learning of the decoherence time formula. In the subsequent training procedures, we will carefully select samples to enhance the distinguishability of the descriptors and their transferability. For simplicity, we use a median error of 0.05 as the threshold. We sort the 323 samples whose FSSH errors are below this threshold in ascending order, then select every 20th sample. In total, we select 16 samples from the STD-2 model, which include 9 samples with the initial WP positioned on the lower PES and 7 samples with the initial WP situated on the upper PES. To enhance the transferability, we further consider a small portion of samples with FSSH errors below 0.06 from the STD-1 model. Namely, two samples have their initial WPs positioned on the lower PES, while the other two samples have their initial WPs located on the upper PES. To ensure consistency across dimensions, we ultimately chose four samples from the 120 samples in the one-dimensional two-level model base MB0 (see Figure S1), whose initial WPs are located on the lower PES. Therefore, the final training set comprises 24 samples, whose detailed parameters are given in Tables S1-S3.

In truth, the direct application of the descriptor space ($\Phi_2$) generated by the basic terms ($E_{kin}$ and $\Delta E$)[53] to delineate high-dimensional problems yields unsatisfactory outcomes. Based on the existing diversity within the training set, we attribute the issue to the characterization capability of the feature input. Inspired by the traditional eq 3, we chose to incorporate a new decoherence correction element and adopt $E_{kin}^{NAC}$, which is calculated from the component of nuclear momenta along the direction of the NAC and naturally reduces to the original $E_{kin}$ for one-dimensional cases. Starting from the original descriptor space $\Phi_0^{NAC} = [E_{kin}^{NAC}, \Delta E]$, we can iteratively obtain the second-order descriptor space $\Phi_2^{NAC}$, which has 371 descriptors in total. Then, the error function for a given decoherence time formula in the training set is calculated using

$$\varepsilon_{train} = \sqrt{\frac{1}{N_{sample}N_{channel}} \sum_{i=1}^{N_{sample}} \sum_{j=1}^{N_{channel}} \left(P_{ij}^{SH} - P_{ij}^{DVR}\right)^2}, \qquad (20)$$

where $N_{channel}$ is uniformly set to 8 for both one- and two-dimensional samples. $N_{sample}$ represents the number of samples in the training set, and its value is 24.

In Figure S2, we show the error distribution of descriptors up to $\Phi_2^{NAC}$. The most remarkable descriptor is located at the top of Table S4, exhibiting an error of 0.010, while the errors of all other descriptors surpass 0.014. The complete definition of the decoherence time formula based on this descriptor is presented as

$$\tau_{ai} = C_0 + C_1 \frac{\Delta E_i}{E_{kin}^{NAC}(E_{kin}^{NAC} - \Delta E_i)} \qquad (21)$$

with $C_0 = 1.0\times10^5$ and $C_1 = 2.0\times10^1$. To evaluate the performance of the above formula, we conduct an error distribution analysis on a comprehensive dataset comprising 1,732 two-dimensional samples, which correspond to diverse initial conditions within the STD models. As shown in Figure 3, 75% of the data points exhibit errors below 0.01 with decoherence-corrected FSSH by eq 21, whereas this percentage is only 37% for the conventional FSSH without decoherence. Furthermore, the error distribution of surface hopping with eq 21 is significantly narrower than that of FSSH, which also underscores the importance of decoherence correction in high-dimensional systems.

To make a quantitative analysis of the decoherence correction with eq 21, we calculate the average population error by

$$\varepsilon_{test}(k_0) = \sqrt{\frac{1}{N_y N_\theta N_{channel}} \sum_{y_0}^{N_y} \sum_{\theta_0}^{N_\theta} \sum_{j=1}^{N_{channel}} (P_{ij}^{SH} - P_{ij}^{DVR})^2} \qquad (22)$$

with $N_{channel} = 8$. Here, $N_y$ and $N_\theta$ denote, respectively, the different possible cases for the initial WP's position and the possible angels between its central momentum and the positive *x*-axis. In the STD-1 model, the values of $N_y$ and $N_\theta$ are 8 and 4, respectively. In the STD-2 model, $N_y$ is equal to 6 and $N_\theta$ is 3. Similarly, in the STD-3 model, $N_y$ is maintained at 6 while $N_\theta$ remains 4. In the full momentum range of the STD-1 model, the error of FSSH remains significant. In particular, it is approximately 0.06 when the initial WPs are located on the upper state with $k_0 = 8$ (see Figure 4A). By introducing the decoherence correction with eq 3, the error can be reduced to approximately half of that observed in the FSSH method, although it remains around 0.02. Surface hopping with our new decoherence time formula (eq 21) and the BCSH approach, can further diminish the error by almost fifty percent. And this reduction is particularly remarkable when the initial WPs are on the lower state, with associated errors for both methods dropping below 0.01 (refer to Figure 4B).

In comparison, the performance of the methods in the STD-2 model exhibit distinct behaviors compared to those in the STD-1 model. Considering the error curve of the FSSH method as a benchmark, it becomes increasingly apparent that the effectiveness of the BCSH method in mitigating errors diminishes progressively as the initial momentum escalates (see Figures 4C and 4D). It is rational that the WP will cease to reflect, causing the BCSH method to revert to the FSSH method as the initial momentum increases and surpasses the barrier. However, the error level reaching up to 0.04 in the high momentum region indicates the presence of an additional decoherence mechanism, which cannot be attributed solely to the conventional reflection of the WP due to insufficient energy. In this instance, the approaches based on decoherence time formulas, which implement decoherence corrections throughout the entire dynamical process, effectively alleviate errors. As illustrated in Figures 4C and 4D, the results indicate that FSSH with eq 3 outperforms BCSH. Encouragingly, our new decoherence correction with eq 21 further significantly reduces the error to approximately 0.02 or less, attributable to the suitable decoherence intensity applied across the entire momentum range.

In addition to the STD-1 and STD-2 models, which exhibit significant errors within the FSSH method, the STD-3 model, known for its reduced errors, can also be utilized as a test system to assess the stability of the new decoherence time formula under conditions of weak decoherence. As illustrated in Figures 4E and 4F, the outcomes of the FSSH method in this model are already exceptionally precise, demonstrating an error of less than 0.01. Following the incorporation of decoherence effect, FSSH with eq 3 or eq 21 maintains results with an accuracy comparable to that of FSSH without any decoherence correction. Overall, within the context of the aforementioned two-dimensional systems, our new decoherence correction with eq 21 exhibits consistent and outstanding performance, whereas FSSH with decoherence correction through eq 3 presents more moderate results. In comparison, the BCSH method reveals significant variability, showcasing superior performance in specific two-dimensional scenarios.

In fact, the decoherence time formula (eq 21) we ultimately chose is screened through data within the training set, which exclusively includes channel populations at the final snapshot. Consequently, the time-dependent population variations and spatial occupancy distributions generated by FSSH with eq 21 require further validation to confirm their reliability. In Figure 5A, we illustrate the temporal evolution of the population on the upper state for the STD-1 model, where the initial WP originates from the upper surface with $x_0 = -4.0$, $y_0 = 0.0$, $k_0 = 10.0$, and $\theta_0 = 0°$. It is evident that FSSH with eq 21 and BCSH accurately depict the time evolution of the population, whereas the results from FSSH with eq 3 and FSSH diverge considerably from the quantum references for $t > 800$ au. Subsequently, in Figure 5B, we show the time evolution of the population on the upper surface for the STD-2 model with the initial case of $x_0 = -8.0$, $y_0 = 0.0$, $k_0 = 6.0$, and $\theta_0 = 0°$. In contrast, only our new decoherence correction with eq 21 preserves accuracy, while the other three methods show considerable deviations from the DVR results. Compared with existing methods, the FSSH approach incorporating the decoherence time formula (eq 21) demonstrates significant improvements in reproducing the time evolution of populations.

Given our singular focus on understanding channel populations at the conclusion of the dynamics, without considering the spatial distribution during intermediate processes, the ability to accurately convey spatial information will undoubtedly affirm the reliability of FSSH with eq 21. In Figures 6 and 7, we show the spatial distributions of the lower state population obtained from DVR, FSSH, and FSSH with eq 21 at the corresponding time steps under the conditions illustrated in Figures 5A and 5B, respectively. The results concerning the corresponding upper state components are provided in Figures S7 and S8 of the Supplementary Information (SI). In Figures 6A, 6D, and 6G for the STD-1 model, at $t = 1000$ au, the initial WP on the upper surface arrives at the first interaction region, giving rise to two branches on the lower surface.

Subsequently, the WPs continue their propagation and reach the second turning point, where an additional two WPs are generated on the lower surface (refer to Figures 6B, 6E, and 6H). Ultimately, the WPs are reflected and completely separated, resulting in one primary part on the upper surface (see Figures S7C, S7F, S7I) and four distinct segments on the lower surface (see Figures 6C, 6F, and 6I). At three pivotal moments in Figure 6, FSSH with eq 21 yields a population distribution that closely aligns with the DVR results, whereas FSSH shows considerable discrepancies.

Next, we examine the STD-2 model under the conditions depicted in Figure 5B. As the initial WP of the upper surface enters the interaction region from the negative direction of the $x$-axis, a new WP component is generated on the lower surface at $t = 1000$ au. In fact, the interaction region manifests as a ring-shaped area of a certain width. Subsequently, the WP of the lower surface first enters the right interaction zone, generating new WP branches (see Figures 7B, 7E, and 7H). Later, at $t = 2900$ au, the WP on the upper surface re-enters the interaction region, undergoes reflection and produces additional WP branches, while the WPs on the lower surface exit the interaction region, becoming distinctly separated from one another. As shown in Figures 7 and S8, during the intermediate to later phases of dynamic evolution, the morphology of the population distribution in the FSSH within the STD-2 model exhibits notable discrepancies when compared to the exact solution. In comparison, FSSH with eq 21 accurately reproduces both the shape and intensity of the exact quantum dynamics, thereby illustrating its efficacy in managing branching within the context of re-entering the interaction zone on multiple occasions.

Besides the excellent performance in describing final channel populations, the time evolution of population, and spatial population distribution for two-dimensional systems, we further test thousands of one-dimensional multilevel systems based on FSSH with eq 21. As demonstrated in Figures S9A-S9D, FSSH with eq 21 excels in replicating the results of the BCSH approach, both of which exhibit the lowest errors across four distinct model bases: the two-level model base MB0, the three-level model base MB1, the four-level model base MB2, and the model base MB3 which is characterized by strong repulsive potentials. Furthermore, we examine the efficacy of FSSH with eq 21 across six standard models, namely the simple avoided crossing (SAC), the dual avoided crossing (DAC), the extended coupling with reflection (ECR), the dumbbell geometry (DBG), the double-arch geometry (DAG), and the dual Rosen-Zener-Demkov noncrossing (DRN) models which have consistently served as benchmark systems (see Figures S10-S15). In contrast to the considerable discrepancies noted with the traditional FSSH, the results derived from FSSH with eq 21 correspond precisely with the exact solution across six standard models.

Beyond the aforementioned investigations, we further conduct a systematic comparative analysis to assess the transferability of previous descriptors, $\max(E_{kin}, \Delta E_i)/(E_{kin} - \Delta E_i)$ and $\Delta E_i/[E_{kin}(E_{kin} - \Delta E_i)]$. Then, we replace $E_{kin}$ with $E_{kin}^{NAC}$ while retaining the original parameters trained on dataset 1, and directly apply the above formulas to two-dimensional systems (STD-1, STD-2, and STD-3 models). Detailed parameters are provided in Table S5, and the corresponding results are shown in Figure S16. Notably, benchmarking against conventional FSSH trajectories reveals that both descriptors, when parameterized using dataset 1, yield significantly lower average population errors. This highlights the crucial role of kinetic energy projection along NAC vectors in high-dimensional scattering systems.

Generally speaking, the transferability of descriptor-derived parameters exhibits a strong positive correlation with the diversity of training datasets. In fact, descriptor $\Delta E_i / [E_{kin}^{NAC}(E_{kin}^{NAC} - \Delta E_i)]$ labeled as $\beta$ with parameters derived from dataset 2, has a better performance than the same descriptor with parameters which are obtained from dataset 1. This result is evident in Figure S16, where the red solid dots consistently appear below the light red hollow circles. However, the descriptor $\max(E_{kin}^{NAC}, \Delta E_i) / (E_{kin}^{NAC} - \Delta E_i)$ labeled as $\alpha$, when optimized using a hybrid dataset 2, demonstrates reduced accuracy across most two-dimensional test cases compared to the same descriptor whose parameters are optimized using dataset 1. As shown in Figure S16, the blue solid points are predominantly located above the light blue hollow circles.

To investigate the decoherence mechanisms embedded in the FSSH framework, we plot the phase diagrams of different decoherence time formulas. For descriptor $\max(E_{kin}^{NAC}, \Delta E_i) / (E_{kin}^{NAC} - \Delta E_i)$, the formula constructed using parameters from a mixed dataset 2 lacks the pronounced decoherence region in the lower-left corner, which is present in the formula based on parameters from dataset 1 (see Figures 8A and 8C). The red area in the lower-left corner corresponds to cases where the energy difference is negative and the kinetic energy component is small. Taking the two-level system as an example, this scenario represents the upper state being active, the lower state being nonactive, and the upper state undergoing WP reflection. Encouragingly, for the latest descriptor $\Delta E_i / [E_{kin}^{NAC}(E_{kin}^{NAC} - \Delta E_i)]$ obtained to date, parameters derived from different training sets affect only the width of the decoherence region, without altering its existence (see Figure 8E). Furthermore, under the STD-1 model, we select a specific initial condition and identify regions where the decoherence time in the actual dynamics is less than 10 (see Figures 8B, 8D, and 8F). In the STD-1 model, with an evolution step size of 0.2 and a total evolution duration of 2500, we define instances where the decoherence time falls below 10 as indicative of stronger decoherence effects. The behavior of the decoherence region in the lower-left corner aligns with the corresponding phase diagrams and more details of time-dependent population are showed in Figure S17.

In summary, we have found a more general decoherence time formula based on the machine learning assisted approach, which shows great performance across 56840 one- and two-dimensional samples. In our previous study, $\Delta E_i$ was employed in place of $|\Delta E_i|$ to validate the significance of the energy difference symbol. Building upon this foundation, we further utilize $E_{kin}^{NAC}$ in place of $E_{kin}$, indicating that, when addressing high-dimensional complex scenarios, the component of $E_{kin}$ in the NAC direction serves as a more appropriate descriptor of the decoherence effect. In the quest for representative samples, we take the comparative performance of the BCSH and FSSH methodologies as references. During this exploration, we also observe that the BCSH method exhibits deficiencies when applied to two-dimensional systems such as STD-2 model, indicating that its corresponding reflection criteria require further refinement. In one-dimensional systems, the motion of the WP is limited to two directions: forward and backward, thereby simplifying the identification of WP reflections on the PES. In contrast, in high-dimensional systems, WP motion becomes significantly more complex, making the definitions of reflection and separation increasingly ambiguous. Naturally, confirming that the reflection of high-dimensional WPs ultimately return to the NAC direction will require further verification in additional high-dimensional systems.

Finally, there remain several points that merit further discussion. (1) We are all aware that the current phase correction algorithms remain contentious in their application to high-dimensional systems. However, in practice, due to the absence of more rigorous derivations, the surface hopping methods employed in this study continue to incorporate phase correction. Thus, apart from decoherence, the meticulous derivation of phase correction represents a significant avenue in the exploration of high-dimensional systems. (2) Given the presence of $\Delta E_i$ as the initial descriptor, $\tau_{ai}$ produced by the subsequent descriptor with the corresponding parameters may indeed be negative. In practical dynamics simulations, we directly reduce the wave function of the nonactive state to zero upon detecting that $\tau_{ai}$ is negative, rather than employing eq 1. Indeed, we use the equation of $\tau_{ai} = \max(0, \tau_{ai})$ when it comes to eq 21. Furthermore, we may consider FSSH with eq 21 as a fusion of the classical decoherence time formula, eq 3, which corresponds to long-range decoherence effects, and BCSH, which includes instantaneous decoherence corrections. (3) In the original formula for decoherence time, as presented in eq 3, the decoherent direction corresponds to the NAC direction in regions characterized by strong interactions, while it signifies the vibrational momentum in other areas. For convenience, we consistently adopt the NAC as the decoherence direction throughout the dynamical region examined in this study. Besides, the component of total kinetic energy aligned with the force direction may serve as a preliminary descriptor for future research. (4) The methodology of using machine learning to derive decoherence time formulas has demonstrated remarkable universality across high-dimensional systems. This approach uses key physical quantities from real systems as initial descriptors and subsequently generates a variety of decoherence time expressions through an iterative process. Beyond scattering systems, we are currently extending its application to spin-boson models. (5) As shown in Figures 8A and 8C, we clarify why parameters fitted to richly sampled training sets exhibit reduced transferability for descriptor $\max(E_{kin}^{NAC}, \Delta E_i)/(E_{kin}^{NAC} - \Delta E_i)$. However, the relative merits of the decoherence mechanisms for the two distinct descriptors remain incompletely understood (see Figures 8C and 8E). Although current research suggests that descriptor $\Delta E_i / [E_{kin}^{NAC}(E_{kin}^{NAC} - \Delta E_i)]$ demonstrates superior universality. To gain more systematic insights, we also attempt to map regions of strong decoherence using a series of alternative decoherence time formulas. These studies are currently underway.

AUTHOR INFORMATION

**Corresponding Author**

*Email: ljwang@zju.edu.cn

**ORCID**

Cancan Shao: 0000-0003-0552-894X

Linjun Wang: 0000-0002-6169-7687

**Author Contributions**

The manuscript was written through contributions of all authors. All authors have given approval to the final version of the manuscript.

**Notes**

The authors declare no competing financial interests.

ACKNOWLEDGMENT

We acknowledge the support from the National Natural Science Foundation of China (Grant 22273082), and the High-Performance Computing Center in Department of Chemistry, Zhejiang University.

ASSOCIATED CONTENT

**Supporting Information**

Computational details, screening of descriptors, additional results of two-dimensional models, additional results of the four one-dimensional model bases and the six standard scattering models, and comparison of performance for two representative descriptors.

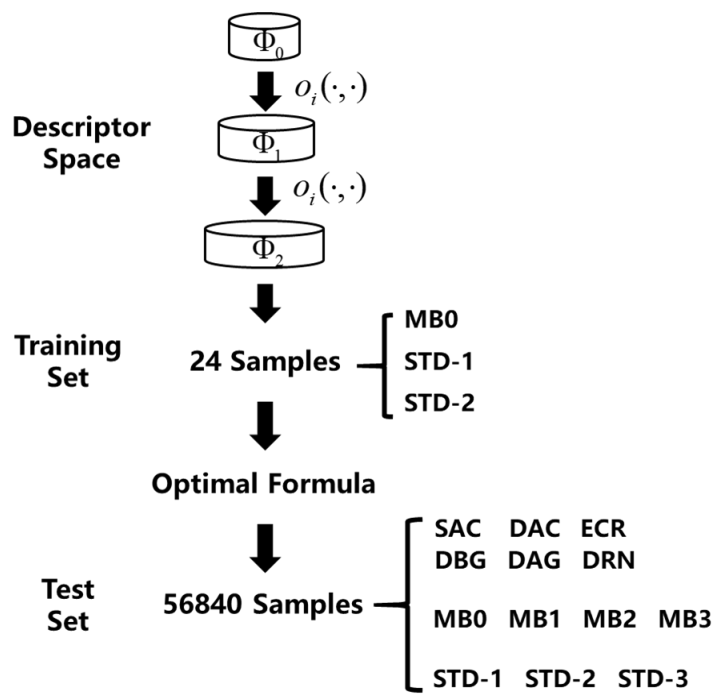

**Figure 1.** Schematic diagram of the screening processes for decoherence time formulas. The training set consists of a total number of 24 representative samples, including 16 samples from the STD-2 models, 4 samples from the STD-1 models, and 4 samples from the two-level model base MB0. The test set contains a total number of 56840 one- and two-dimensional samples.

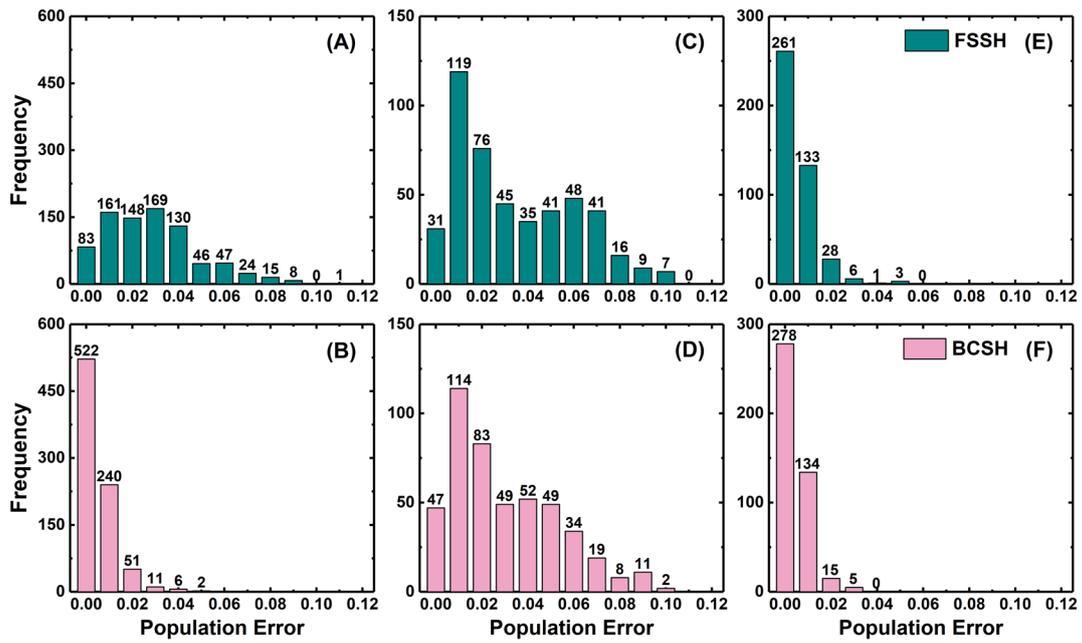

**Figure 2.** Population error distribution obtained by FSSH for the (A) STD-1, (C) STD-2, and (E) STD-3 samples and that by BCSH for the (B) STD-1, (D) STD-2, and (F) STD-3 samples.

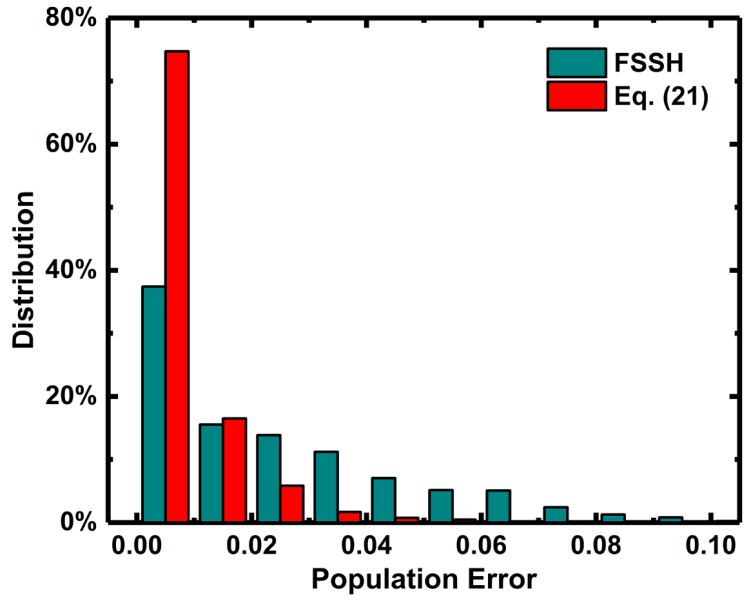

**Figure 3.** Population error distribution for all the STD samples obtained by FSSH and FSSH with the decoherence time calculated by eq 21. 75% of the data points exhibit errors below 0.01 when employing eq 21 as the decoherence time formula, whereas this percentage diminishes to 37% for the standard FSSH method.

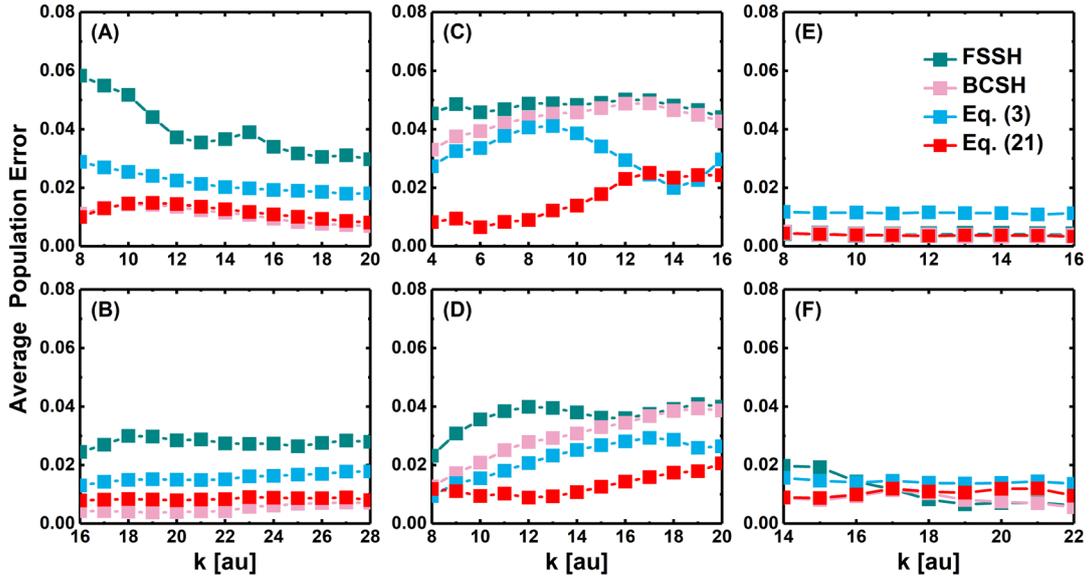

**Figure 4.** Average population error of the STD-1 model initialized from the (A) upper and (B) lower surface, the STD-2 model initialized from the (C) upper and (D) lower surface, and the STD-3 model initialized from the (E) upper and (F) lower surface. The results of FSSH, BCSH, and FSSH with the decoherence time calculated by eqs 3 and 21 are represented by dark cyan, pink, blue, and red solid squares, respectively.

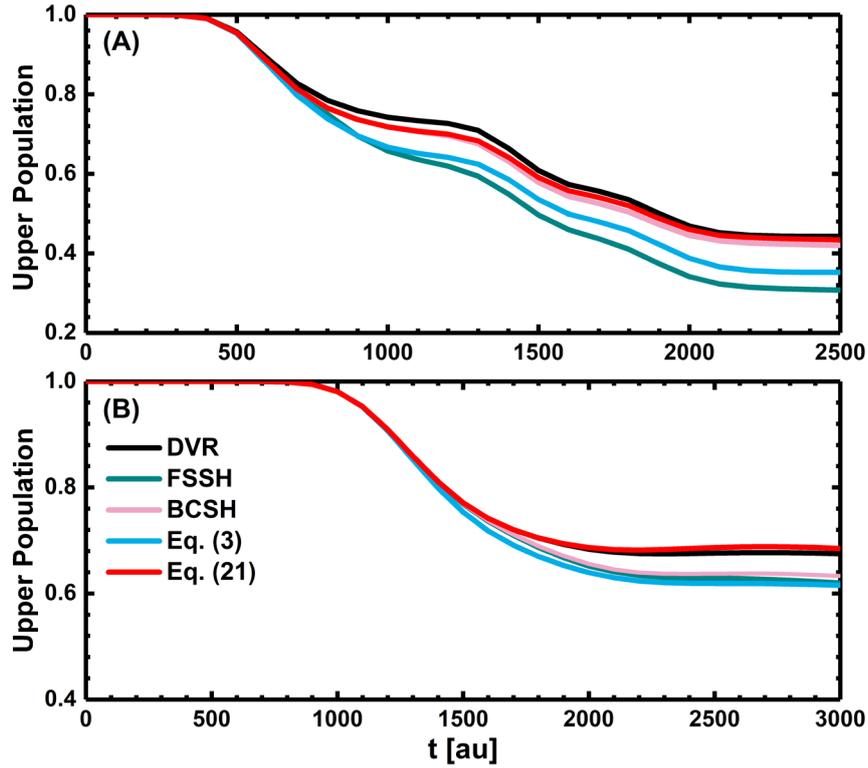

**Figure 5.** Time-dependent population on the upper adiabatic PES for (A) the STD-1 model and (B) the STD-2 model. In (A), the initial WP is situated on the upper PES at $x_0 = -4.0$, $y_0 = 0.0$, $k_0 = 10.0$, and $\theta_0 = 0°$. In (B), the initial WP is positioned on the upper PES at $x_0 = -8.0$, $y_0 = 0.0$, $k_0 = 6.0$, and $\theta_0 = 0°$. The black solid lines represent the exact quantum dynamics with DVR. The results of FSSH, BCSH, and FSSH with the decoherence time calculated by eqs 3 and 21 are depicted by dark cyan, pink, blue, and red solid lines, respectively.

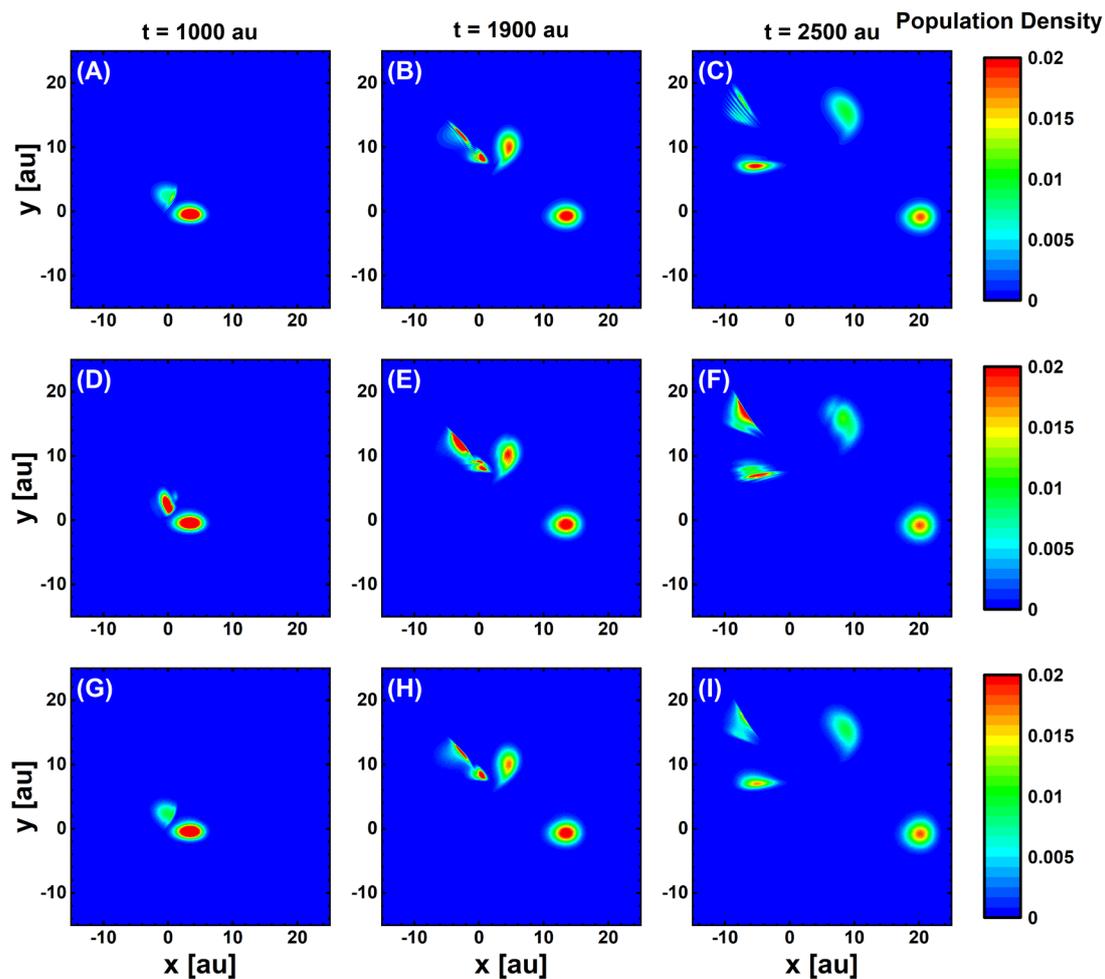

**Figure 6.** Spatial distribution of population on the lower adiabatic PES of the STD-1 model obtained by (A-C) exact quantum dynamics with DVR, (D-F) FSSH, and (G-I) FSSH incorporating eq 21. The initial conditions are identical to those depicted in Figure 5A. (A, D, and G), (B, E, and H), and (C, F, and I) correspond to $t$ = 1000, 1900, and 2500 au, respectively.

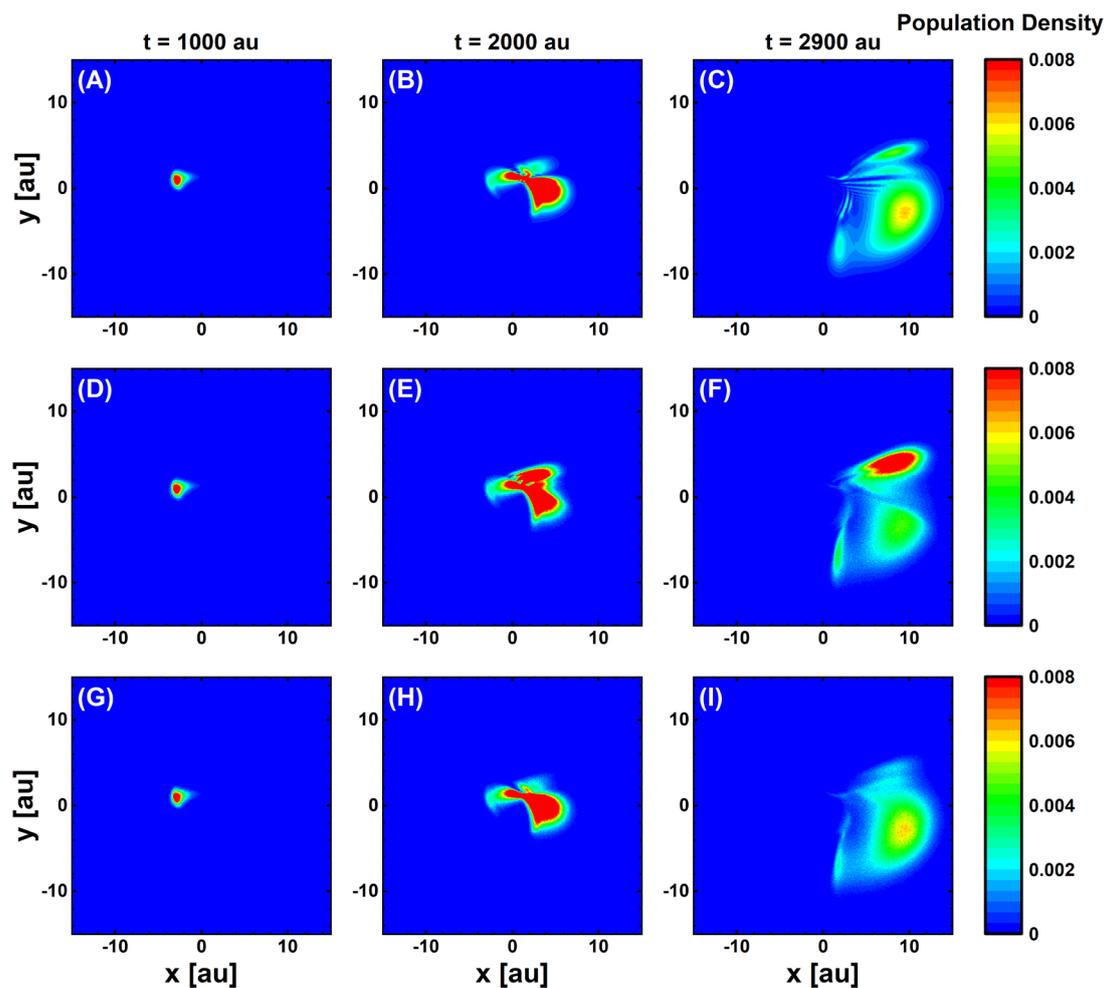

**Figure 7.** Spatial distribution of population on the lower adiabatic PES of the STD-2 model obtained by (A-C) exact quantum dynamics with DVR, (D-F) FSSH, and (G-I) FSSH incorporating eq 21. The initial conditions are identical to those depicted in Figure 5B. (A, D, and G), (B, E, and H), and (C, F, and I) correspond to $t$ = 1000, 2000, and 2900 au, respectively.

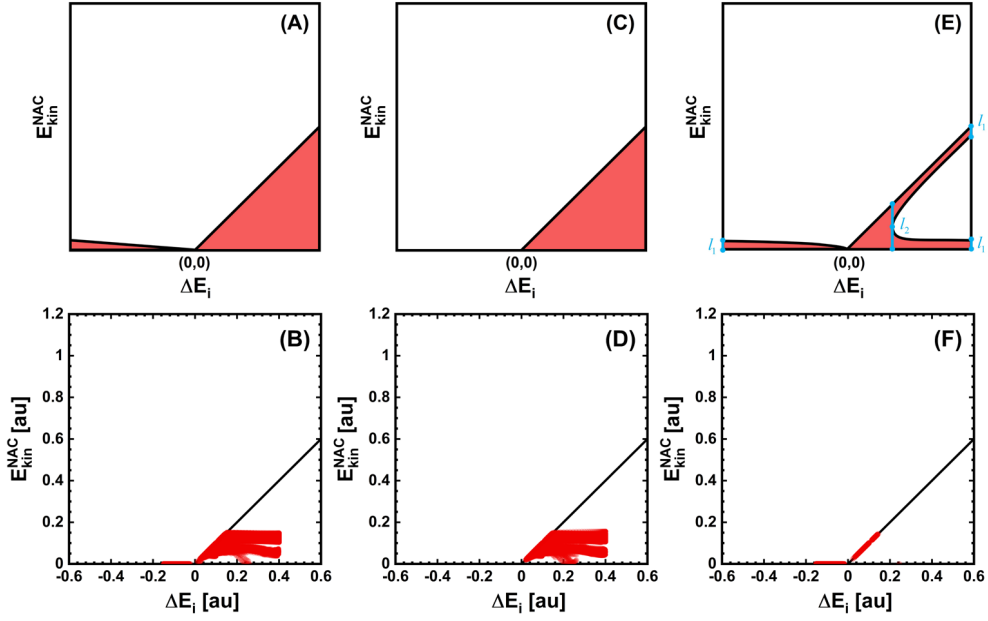

**Figure 8.** Distribution of the decoherence time in the two-dimensional space of $\Delta E_i$ and $E_{kin}^{NAC}$. The regions highlighted in red indicate that the decoherence time is less than 10. The decoherence time formula adopted in (A) and (B) correspond to the descriptor $\max(E_{kin}^{NAC}, \Delta E_i)/(E_{kin}^{NAC} - \Delta E_i)$ with $[C_0 = 2.5 \times 10^{-3}, C_1 = 2.0 \times 10^5]$, while (C) and (D) utilize the same descriptor but with another set of parameters $[C_0 = 5.0 \times 10^2, C_1 = 6.4 \times 10^2]$. In (E) and (F), the decoherence time formula is eq 21, which is based on the descriptor $\Delta E_i / [E_{kin}^{NAC}(E_{kin}^{NAC} - \Delta E_i)]$ and uses the parameters $[C_0 = 1.0 \times 10^5, C_1 = 2.0 \times 10^5]$. In (E), $\lim_{\Delta E_i \to \infty} l_1 = C_1/(C_0 - 10)$ and $l_2 = 4C_1/(C_0 - 10)$.